\documentclass[prd,aps,showpacs,%11pt,%showpacs,
nofootinbib,%
twocolumn%,superscriptaddress,
]{revtex4-1}

\usepackage{graphicx}
\usepackage{amsmath}
\usepackage{amssymb}
\usepackage{hyperref} 

\newcommand{\be}{\begin{equation}}
\newcommand{\ee}{\end{equation}}
\newcommand{\bea}{\begin{eqnarray}}
\newcommand{\eea}{\end{eqnarray}}
\newcommand{\nn}{\nonumber}

\newcommand{\MKK}{M_{\rm KK}}

\newcommand{\Tr}{{\rm Tr}\,}

\newcommand{\D}{G_{D}}

\def\PDG{\cite{Agashe:2014kda}}
\def\BPR{\cite{Brunner:2015oqa}}
\def\BR{\cite{1504.05815}}

\begin{document}

\title{%Correlation between flavor asymmetry and $\eta\eta'$~rate%branching ratio
%\\ in holographic glueball decay%\\
%Predicting the $\eta\eta'$~decay rate of scalar glueballs from
Constraints on the $\eta\eta'$ decay rate of a scalar glueball from
gauge/gravity duality
%the\\ Witten-Sakai-Sugimoto model with finite quark masses
}

%\preprint{TUW-15-XX}

\author{Frederic Br\"unner}
\author{Anton Rebhan}
\affiliation{Institut f\"ur Theoretische Physik, Technische Universit\"at Wien,
        Wiedner Hauptstrasse 8-10, A-1040 Vienna, Austria}

\date{\today}

\begin{abstract}
Predictions of glueball decay rates in
the holographic Witten-Sakai-Sugimoto model for low-energy QCD 
can be uniquely extended to include finite quark masses up to
an as yet undetermined parameter in the coupling of glueballs to
the nonanomalous part of the pseudoscalar mass terms.
The assumption of a universal 
coupling of glueballs to mass terms of the full nonet of pseudoscalar mesons
leads to flavor asymmetries in the decay rates of scalar glueballs that agree well
with experimental data for the glueball candidate $f_0(1710)$ 
and implies a vanishing decay rate into
$\eta\eta'$ pairs, for which only upper bounds for the $f_0(1710)$ meson are known at present from experiment.
Relaxing this assumption, the holographic model gives
a tight correlation between the decay rates into pairs of pseudo-Goldstone bosons of
same type and $\eta\eta'$ pairs.
If $\Gamma(G\to KK)/\Gamma(G\to\pi\pi)$ is kept within the range reported currently by the Particle Data Group
for the $f_0(1710)$ meson, the
rate $\Gamma(G\to\eta\eta')/\Gamma(G\to\pi\pi)$ is predicted to be $\lesssim 0.04$.
The corresponding situation for $f_0(1500)$ is also discussed, 
which however is found to be much less
compatible with the interpretation of a largely unmixed glueball.
\end{abstract}

\maketitle

\section{Introduction}

While quantum chromodynamics (QCD) has been established beyond any reasonable doubt
as the fundamental theory of hadrons, one of its most conspicuous
predictions, the existence of
bound states of gluons called glueballs or gluonia \cite{Fritzsch:1972jv,Fritzsch:1975tx,Jaffe:1975fd}
as well as their nature
have still not been settled by experiment
\cite{Bugg:2004xu,Klempt:2007cp,Crede:2008vw,Ochs:2013gi}.
Lattice QCD \cite{Bali:1993fb,Morningstar:1999rf}
predicts the lightest glueball to be a scalar with mass
in the range 1.5-1.8 GeV, with only small effects from
unquenching \cite{Gregory:2012hu}, but predictions from first principles on
the potential mixing of glueballs with scalar quark-antiquark states and the decay pattern
of glueballs are hard to come by. 

The expectation from QCD in the limit of large number of colors ($N_c$) is that
glueballs should be comparatively narrow states, and also that mixing
should be suppressed \cite{Lucini:2012gg}. If this holds true for $N_c=3$ QCD, holographic
gauge/gravity duality might be useful to shed light on the nature of glueballs.
Here a particularly attractive model is due to Sakai and Sugimoto 
\cite{Sakai:2004cn,Sakai:2005yt} who have extended the Witten model \cite{Witten:1998zw}
for nonsupersymmetric and nonconformal low-energy QCD based on D4 branes in
type-IIA supergravity by $N_f\ll N_c$ D8 and anti-D8 branes, which introduce
chiral quarks and allow for a purely geometric realization of chiral symmetry breaking
$\mathrm{U}(N_f)_L\times\mathrm{U}(N_f)_R
\to\mathrm{U}(N_f)_V$.

As reviewed in \cite{Rebhan:2014rxa}, with almost no free parameters
the (massless) Witten-Sakai-Sugimoto model reproduces many features
of low-energy QCD and turns out to work remarkably well even on a quantitative level, although the
model has a Kaluza-Klein mass scale $\MKK$ at the order of glueball masses beyond which the dual theory is
actually a five-dimensional super-Yang-Mills theory (whose extra degrees of freedom are discarded by
consistent truncation). 
Fixing $\MKK$ through the experimental value of the $\rho$ meson mass and varying 
the 't Hooft coupling $\lambda=16.63\ldots12.55$ such that either the pion decay constant (as originally
done in \cite{Sakai:2004cn,Sakai:2005yt})
or the string tension in large-$N_c$ lattice simulations \cite{Bali:2013kia} is matched, the decay rate of
the $\rho$ and the $\omega$ meson into pions is obtained as \BPR\
\bea
&&\Gamma(\rho\to2\pi)/ m_\rho = 0.1535\dots 0.2034,\\
&&\Gamma(\omega\to3\pi)/ m_\omega = 0.0033\ldots 0.0102,
\eea
which matches well with the respective experimental values, 0.191(1) and
0.0097(1) \PDG.
The Witten-Sakai-Sugimoto model might therefore be capable of giving useful information on the decay
patterns of glueballs, in particular if they exist without being mixed too strongly with $q\bar q$ states.

A large number of phenomenological studies assume the lowest scalar glueball
to be responsible for the supernumerary state among the three isoscalar mesons $f_0(1370)$, 
$f_0(1500)$, and $f_0(1710)$, for which the quark model
provides just $u\bar u+d\bar d$ and $s\bar s$,
but there is no agreement whether the glueball is to be found predominantly in
$f_0(1500)$ or $f_0(1710)$ 
\cite{Amsler:1995td,Lee:1999kv,Close:2001ga,Amsler:2004ps,Close:2005vf,Giacosa:2005zt,Albaladejo:2008qa,Mathieu:2008me,Janowski:2011gt,Janowski:2014ppa,Cheng:2015iaa,Close:2015rza,Frere:2015xxa}.
Most of the earlier analysis described $f_0(1500)$ as having a large glueball component with sizable mixing,
while later studies (though not all of them) appear to prefer the interpretation of $f_0(1710)$ 
as a glueball with rather small mixing.

Experimentally, the decay pattern of the isoscalar $f_0(1710)$ is still not known as precisely as that
of $f_0(1500)$. However, it is well known that $f_0(1710)$ decays preferentially into kaons and $\eta$ mesons and much less into
pions, while naively one would expect flavor-blindness in glueball decays.
The flavor asymmetry in the decay of $f_0(1710)$ has been attributed to
the mechanism of ``chiral suppression'' \cite{Carlson:1980kh,Sexton:1995kd,Chanowitz:2005du}
according to which (part of the) decay amplitudes of a scalar glueball should be proportional to
quark masses. However, in view of chiral symmetry breaking and the resulting large constituent quark masses
this argument seems to be questionable
\cite{Frere:2015xxa}.

In extension of our previous work on glueball decay in the Witten-Sakai-Sugimoto model \BPR,
we have recently shown \BR\ that an effectively equivalent mechanism we
called nonchiral enhancement could explain these flavor asymmetries.
The assumption of a universal 
coupling of glueballs to anomalous and nonanomalous mass terms of the full nonet of pseudoscalar mesons
leads to a remarkably good agreement
with experimental data for the glueball candidate $f_0(1710)$ while 
implying a vanishing decay rate into
$\eta\eta'$ pairs, for which only upper bounds for the $f_0(1710)$ meson are at present available from experiment.\footnote{An
enhancement of the glueball decay rate into pseudoscalars according to the mass of the latter
(instead of a complete suppression in the chiral limit) was also found in the model of Ref.~\cite{Ellis:1984jv},
albeit just large enough to compensate approximately for kinematic suppression.}

In this paper we relax this assumption and study the predictions of
the Witten-Sakai-Sugimoto model with finite quark masses when the as yet undetermined
parameter in the glueball coupling to the nonanomalous mass terms is kept free.
This will in particular lead to constraints on the $\eta\eta'$ decay rate of
$f_0(1710)$ when interpreted as a nearly unmixed glueball state.

We shall also confirm our conclusion in \BPR\ that the glueball candidate $f_0(1500)$ 
is disfavored by the holographic model, although the mass of the lowest predominantly dilatonic mode that
we identify with the lightest scalar glueball of QCD\footnote{In the first attempt to calculate the decay rate of the lightest glueball in \cite{Hashimoto:2007ze} the lightest mode of the spectrum of glueballs
obtained originally in
\cite{Constable:1999gb,Brower:2000rp} was identified with the lightest glueball, although when $\MKK$ is
fixed by the experimental value of the $\rho$ meson mass it comes out at 855 MeV and thus much too light
compared to lattice results. In \BPR\ we have argued
that the lowest-lying mode, which involves an ``exotic'' \cite{Constable:1999gb} polarization along the compactification direction,
should be discarded and replaced by the next-highest, predominantly dilatonic mode.}
is 1487 MeV and thus very close to $f_0(1500)$.
% A first attempt to calculate the decay rate of the lightest glueball was made in \cite{Hashimoto:2007ze},
% using the lightest mode of the spectrum of glueballs
% obtained originally in
% \cite{Constable:1999gb,Brower:2000rp}. Revisiting these calculations \BPR\ we have argued
% that the lowest-lying mode, which involves an ``exotic'' \cite{Constable:1999gb} polarization along the compactification direction,
% should be discarded and replaced by the next-highest, predominantly dilatonic mode.
% Its mass of 1487 MeV is close to expectations from lattice QCD,
However, the resulting decay pattern \BPR\ does not fit well to that of the $f_0(1500)$ meson
when interpreted as a pure glueball.
The decay into two pions is underestimated by a factor of 2, and the rate into four pions, which is the
dominant decay mode of $f_0(1500)$, is too small by almost an order of magnitude.

This leaves the $f_0(1710)$ as a candidate for a nearly unmixed glueball.
After all, the mass of $f_0(1710)$ is only 16\% heavier, and we cannot expect the holographic
model to be more accurate than some 10-30\%.
Indeed, the decay rate of $f_0(1710)$ into two pions seems to be of roughly the right magnitude, 
while the significantly higher rates into kaons and eta mesons
may be attributed to the nonchiral enhancement effect found in \BR\
under the assumption of a universal coupling of glueballs to pseudoscalar mass terms.

Before studying the effect of finite quark masses on glueball decay in the Witten-Sakai-Sugimoto model
in maximal generality, we discuss the mass term arising from the U(1)$_A$ anomaly and its interplay
with finite quark masses.

\section{Witten-Veneziano mass and pseudoscalar mixing}

In the Witten-Sakai-Sugimoto model, the U(1)$_A$ flavor symmetry is broken by anomaly contributions of
order $1/N_c$, which give rise to a 
Witten-Veneziano \cite{Witten:1979vv,Veneziano:1979ec} mass term for the singlet $\eta_0$ pseudoscalar
that is local with respect to the effective 3+1-dimensional
boundary theory,
\be\label{LmWV}
\mathcal L_m^{WV}=-\frac12 m_0^2 \eta_0^2(x),
\ee
where $\eta_0$ is obtained by an integration over the bulk coordinate $z$,
\be
\eta_0(x)=\frac{f_\pi}{\sqrt{2N_f}}{\rm Tr} \int dz A_z(z,x).
\ee
The Witten-Veneziano mechanism relates $m_0^2$ to the topological susceptibility. It has been calculated
by Sakai and Sugimoto in their model (following earlier work by \cite{Armoni:2004dc,Barbon:2004dq}) with the result \cite{Sakai:2004cn}
\be\label{mWV2}
m_{0}^2=\frac{N_f}{27\pi^2 N_c}\lambda^2\MKK^2.
\ee
For $N_f=N_c=3$, and the choice of parameters considered
by us in \cite{Brunner:2015oqa,1504.05815}, namely $\MKK=949$ MeV, and $\lambda$ varied from 12.55 to 16.63, one finds
$m_{0}=730$ - $967$ MeV.

The massless Witten-Sakai-Sugimoto model can in principle be deformed to include
mass terms for the entire pseudoscalar nonet by either worldsheet instantons \cite{Aharony:2008an,Hashimoto:2008sr}
or nonnormalizable modes of bifundamental fields corresponding to open-string tachyons \cite{0708.2839,Dhar:2008um,McNees:2008km,Niarchos:2010ki}. These scenarios have only
been demonstrated on a qualitative level, but they all agree in the form of those mass terms,
%which are similarly nonlocal with respect to the bulk and of the form
\bea\label{calM}
&&\mathcal L_m^{\mathcal M} \propto\int d^4x \,\Tr\left(\mathcal M\,U(x)+h.c.\right),\\
%\be\label{UPi}
&&U(x)=\mathrm P\,\exp i\int_{-\infty}^\infty dz A_z(z,x)=e^{i\Pi^a\lambda^a/f_\pi},
\eea
with tunable $\mathcal M={\rm diag}(m_u,m_d,m_s)$, so that Gell--Mann-Oakes-Renner relations are realized.

For simplicity, and because we are not going to include electromagnetic interactions,
we shall keep isospin symmetry and set $m_u=m_d=\hat m$.
The mass terms resulting from $\mathcal M$ then read
\bea\label{LmM}
\mathcal L_m^{\mathcal M}&=&-\frac12 m_\pi^2 \pi_0^2-m_\pi^2 \pi_+\pi_- \nn\\
&&-m_K^2(K_0 \bar K_0 + K_+ K_-) \\
&&-\frac12 m_1^2 \eta_0^2-\frac12 m_8^2 \eta_8^2 -
\frac{2\sqrt2}{3}(m_K^2-m_\pi^2)\eta_0 \eta_8\nn
\eea
with
\bea\label{mpiK}
&&m_\pi^2=2\hat m \mu,\quad m_K^2=(\hat m+m_s)\mu,\\
&&m_1^2=\frac23 m_K^2+\frac13 m_\pi^2,\quad
m_8^2=\frac43 m_K^2-\frac13 m_\pi^2,
\eea
and $\mu$ being the overall scale in (\ref{calM}).

With the addition of mass terms of the form (\ref{calM}), $\eta_0$ and $\eta_8$ are no longer mass eigenstates.
Defining the mass eigenstates in terms of the pseudoscalar mixing angle $\theta_P$ \PDG
\bea
\eta &=& \eta_8 \cos\theta_P - \eta_0 \sin\theta_P\nn\\
\eta'&=& \eta_8 \sin\theta_P + \eta_0 \cos\theta_P,
\eea
one obtains
\bea
m^2_{\eta,\eta'}&=&\frac12 m_0^2+m_K^2\nn\\
&&\mp\sqrt{\frac{m_0^4}{4}-\frac13 m_0^2(m_K^2-m_\pi^2)+(m_K^2-m_\pi^2)^2},\qquad\\
\theta_P&=&\frac12\arctan\frac{2\sqrt2}{1-\frac32 {m_0^2}/({m_K^2-m_\pi^2})}.
\eea

In the absence of the Witten-Veneziano mass term (\ref{LmWV}), one
would find $m_\eta^2=m_\pi^2$ and $m_{\eta'}^2=2m_K^2-m_\pi^2$,
whereas $m_0\to\infty$ with $\theta_P\to0$ corresponds to a decoupling
of $\eta'$ and $m_\eta^2\to m_8^2$.

In the Witten-Sakai-Sugimoto model, $m_0^2$ is given by (\ref{mWV2}),
and our inclusion of quark masses through $\hat m \mu$ and $m_s\mu$
does not add any free parameters, as those can be fixed by the experimental values of $m_\pi$ and $m_K$.
In order to obtain an optimal match of $\hat m$, the average of $m_u$ and $m_d$, 
without isospin
breaking mass contributions from the electromagnetic interactions, we shall
use $m_\pi^2=m_{\pi_0}^2\approx (135 {\rm MeV})^2$ in (\ref{mpiK}), while $m_s$ will be fixed by
\be
m_K^2=\frac12(m_{K_\pm}^2+m_{K_0}^2)-
\frac12(m_{\pi_\pm}^2-m_{\pi_0}^2)\approx (495{\rm MeV})^2,
\ee
corresponding to a ratio $m_s/\hat m\approx 25.9$ slightly below the current-quark
mass ratio $27.5\pm1.0$ of Ref.~\PDG.

This setup for the masses in fact
never reaches the experimental
mass ratio $m_\eta/m_{\eta'}\approx 0.572$ for any values of $m_u,m_d,m_s,m_{0}$
\cite{Georgi:1993jn,Gerard:2004gx}. With our isospin symmetric choice of quark masses a maximum
value of $m_\eta/m_{\eta'}\approx 0.535$ is attained at $m_0\approx 660.6$~MeV and $\theta_P\approx -28^\circ$, which is however
outside the range considered above for the Witten-Sakai-Sugimoto model.
But at this optimal point with respect to 
the ratio of the masses, both $m_\eta$ and $m_{\eta'}$ are about 100 MeV below their
physical values.

\begin{figure}
\includegraphics[width=0.45\textwidth]{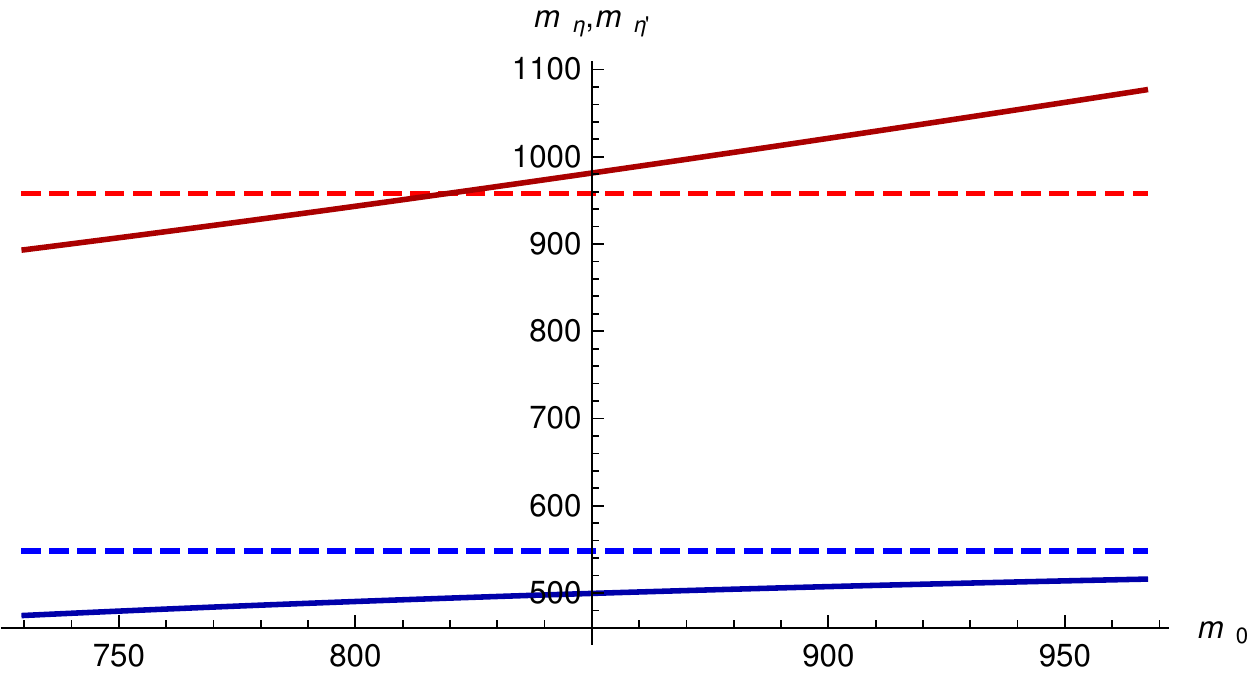}
\includegraphics[width=0.45\textwidth]{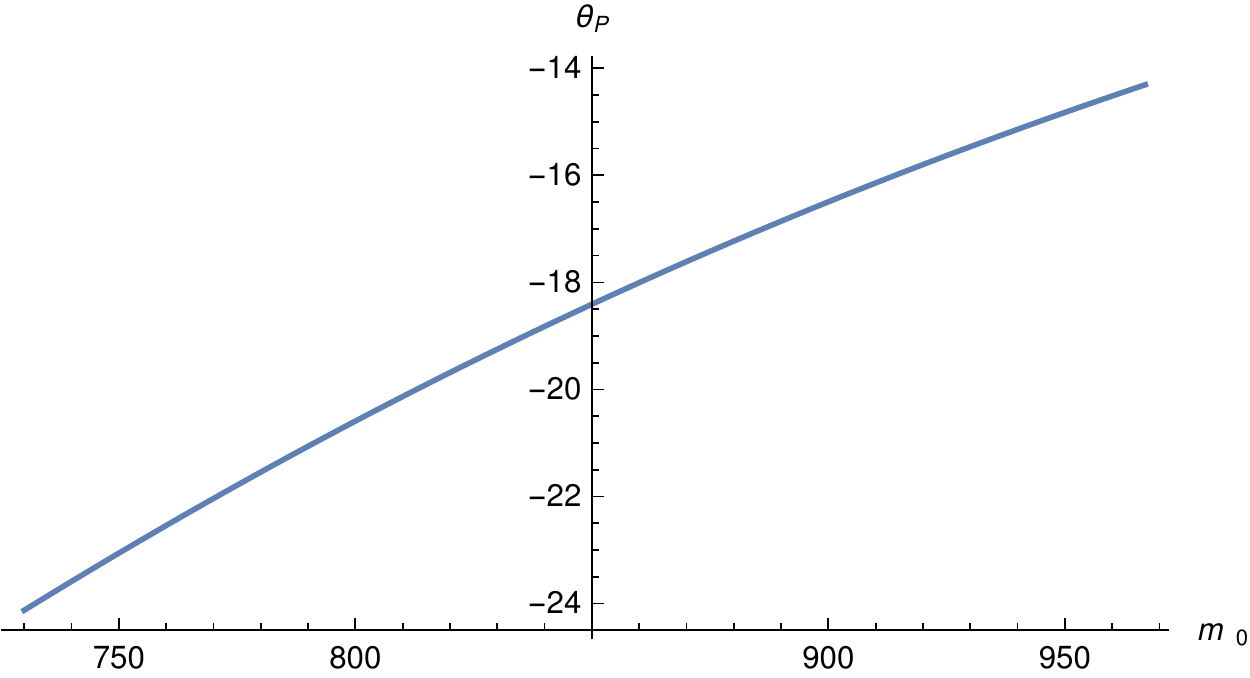}
\caption{Masses of the $\eta$ and $\eta'$ meson (blue and red lines in upper plot) and pseudoscalar mixing angle $\theta_P$ (lower plot)
as a function of the Witten-Veneziano mass $m_0$ in the range obtained for the latter in the Witten-Sakai-Sugimoto model.}
\label{figmetas}
\end{figure}

The range of values for $m_\eta$, $m_{\eta'}$, and $\theta_P$ obtained for $m_0\in(730,967)$MeV
are shown in Fig.~\ref{figmetas}. It turns out that 
just around the center value of this range one can achieve various other
possible optimizations of $m_\eta$ and $m_{\eta'}$: the sum $m_\eta+m_{\eta'}$ matches the experimental
value at $m_0\approx 876$~MeV and $\theta_P\approx -17.4^\circ$; for the sum of squared masses this
is the case at $m_0\approx 852$~MeV and $\theta_P\approx -18.3^\circ$; a least-square value of errors
for the masses is obtained at $m_0\approx 835$~MeV and $\theta_P\approx -19^\circ$.

The total range of $\theta_P$ in Fig.~\ref{figmetas} coincides with most of the range considered
in the phenomenological literature:
From light meson decays values of $\theta_P$ around $-14^\circ$ appear to be favored \cite{Ambrosino:2009sc,Pham:2015ina}, 
while radiative charmonium decay
instead points to $\theta_P\approx -21^\circ$ \cite{Gerard:2004gx,Gerard:2013gya}.
Also the ratio $\Gamma(\eta'\to2\gamma)/\Gamma(\eta\to2\gamma)$ leads to larger values \PDG,
$\theta_P=(-18\pm2)^\circ$, which happens to include the various optimization points of $\eta$ and $\eta'$ masses listed above.

In the following we shall use $m_0=850$~MeV as a central value, and $m_0\in(730,967)$MeV as a range for estimating
a theoretical error bar for our semi-quantitative study of the Witten-Sakai-Sugimoto model. It should of course
be kept in mind that we have no control over subleading order corrections in this model, but our hope is that
errors continue to remain in the 10-30\% range seen in various previous applications \cite{Rebhan:2014rxa}. In fact, if one considers
the values of $m_\eta$ and $m_{\eta'}$ as another prediction of the Witten-Sakai-Sugimoto model,
the errors are, encouragingly, in the $\lesssim10$\% range only.

\section{Glueball couplings and decay rates in the Witten-Sakai-Sugimoto model with quark masses}

For the chiral Witten-Sakai-Sugimoto model the vertices of glueball fields with pseudoscalar and vector mesons
have been worked out in detail in \BPR. The
vertex for an on-shell scalar glueball 
corresponding to the lowest predominantly dilatonic mode $G_D$ with mass $M$
and two pseudoscalar mesons reads
\be\label{LGDchiral}
\mathcal L_{\D\pi\pi}^{\rm chiral}= \frac{1}{2}d_1 \Tr (\partial_\mu\boldsymbol\pi\partial_\nu\boldsymbol\pi)
\left(\eta^{\mu\nu}-\frac{\partial^\mu \partial^\nu}{M^2}\right)\D,
\ee
where $d_1\approx 17.226 \lambda^{-1/2}N_c^{-1}\MKK^{-1}$ and $\boldsymbol\pi=\Pi^a\lambda^a/\sqrt2$ (with
the $\lambda^a$ including the U(1) generator through $a=0,\ldots,8$).

An additional contribution to the
vertex of the U(1) pseudoscalar with $G_D$ arises from the Witten-Veneziano mass term
which has been calculated in \cite{1504.05815} as
\be
\mathcal L_{\D\eta_0\eta_0}^{\rm chiral}=\frac32 d_0 m_{0}^2 \,\eta_0^2 G_D
\ee
with $d_0\approx {17.915}\lambda^{-1/2}N_c^{-1}\MKK^{-1}$.
This coupling follows unambiguously from the original Witten-Sakai-Sugimoto model with massless quarks.

The coupling of $G_D$ to the mass terms of the pseudo-Goldstone bosons induced by quark masses
through tachyon condensation or world-sheet instantons is, however, not known.
Given the similarity of the holographic expressions for the anomalous and the explicit mass terms,
it is however clear that the latter will also have a coupling to the glueball modes which are concentrated around the minimal
value of the radial bulk coordinate.
In \cite{1504.05815} the simplest possibility was explored, namely that of a universal coupling of
the glueball to both kinds of meson mass terms, for which plausibility arguments were given. 
This most symmetric scenario implies that the 
mass terms of all pseudo-Goldstone bosons can be diagonalized simultaneously 
with their glueball couplings so that no $G_D\eta\eta'$ vertex arises. In this paper 
we shall study the consequences of having
an undetermined prefactor $d_m$ in the glueball couplings to the pseudoscalar mass terms arising from
nonzero quark masses, i.e.
\be
\mathcal L_{\D\pi\pi}^{\rm massive}=\frac32 d_m G_D\mathcal L_m^{\mathcal M}
\ee
with $L_m^{\mathcal M}$ given by (\ref{LmM}). The deviation of $d_m$ from $d_0$ will be
denoted by $$d_m\equiv x d_0$$ with a free parameter $x$.

The decay rate of a glueball into two pseudoscalar mesons
%pions or two kaons 
then has a mass dependence
according to
\bea\label{enhancementPP}
&&\Gamma(G_D\to PP)^{\rm massive} %/\Gamma(G_D\to PP)^{\rm chiral}
\nn\\
&&=
\frac{n_P\, d_1^2 M^3}{512\pi}
\left(1-4\frac{m_P^2}{M^2}\right)^{1/2}\left(1+\alpha_P \frac{m_P^2}{M^2}\right)^2
\eea
with
\be\label{alphaD}
\alpha_P=4\left(3 \frac{d_0}{d_1} x - 1\right)\approx 4(3.120\,x-1)
\ee
when $P$ refers to pions ($n_P=3$) or kaons ($n_P=4$).
Note that in the (leading-order) result (\ref{alphaD}) the dependence on $\lambda$ and $\MKK$ has dropped out; it
only depends on the $O(1)$ parameter $x$, which parametrizes the undetermined ratio $d_m/d_0$.

For $\eta$ mesons ($n_P=1$) 
the modification factor $\alpha$ depends on the $\eta$-$\eta'$ mixing (and thus through $m_0$
on $\lambda$ and $\MKK$).
It reads
\be
\alpha_\eta=
4\left(3  \frac{d_0}{d_1}\left[x+\sin^2\theta_P\frac{m_0^2}{m_\eta^2}(1-x)\right] - 1\right).
\ee
(For glueballs heavy enough to be able to decay into two $\eta'$ mesons the corresponding
quantity would involve $\cos^2\theta_P$ in place of $\sin^2\theta_P$.)

With $d_m\not=d_0$, a scalar glueball can also decay into an $\eta$-$\eta'$ pair through
the vertex
\be
\mathcal L_{\D\eta\eta'}^{\rm massive}=
-\frac32 (d_0-d_m)\sin(2\theta_P)m_0^2 \,\D\,\eta\,\eta'
\ee
with rate
\bea
&&\Gamma(\D\to\eta\eta')=\frac{|\mathbf p|}{8\pi M^2}
\left( \frac32 (d_0-d_m)\sin(2\theta_P)m_0^2 \right)^2,\nn\\
&&|\mathbf p|=\frac{\sqrt{[M^2-(m_\eta+m_{\eta'})^2][M^2-(m_\eta-m_{\eta'})^2]}}{2M}.
\eea

\begin{figure}
\includegraphics[width=0.45\textwidth]{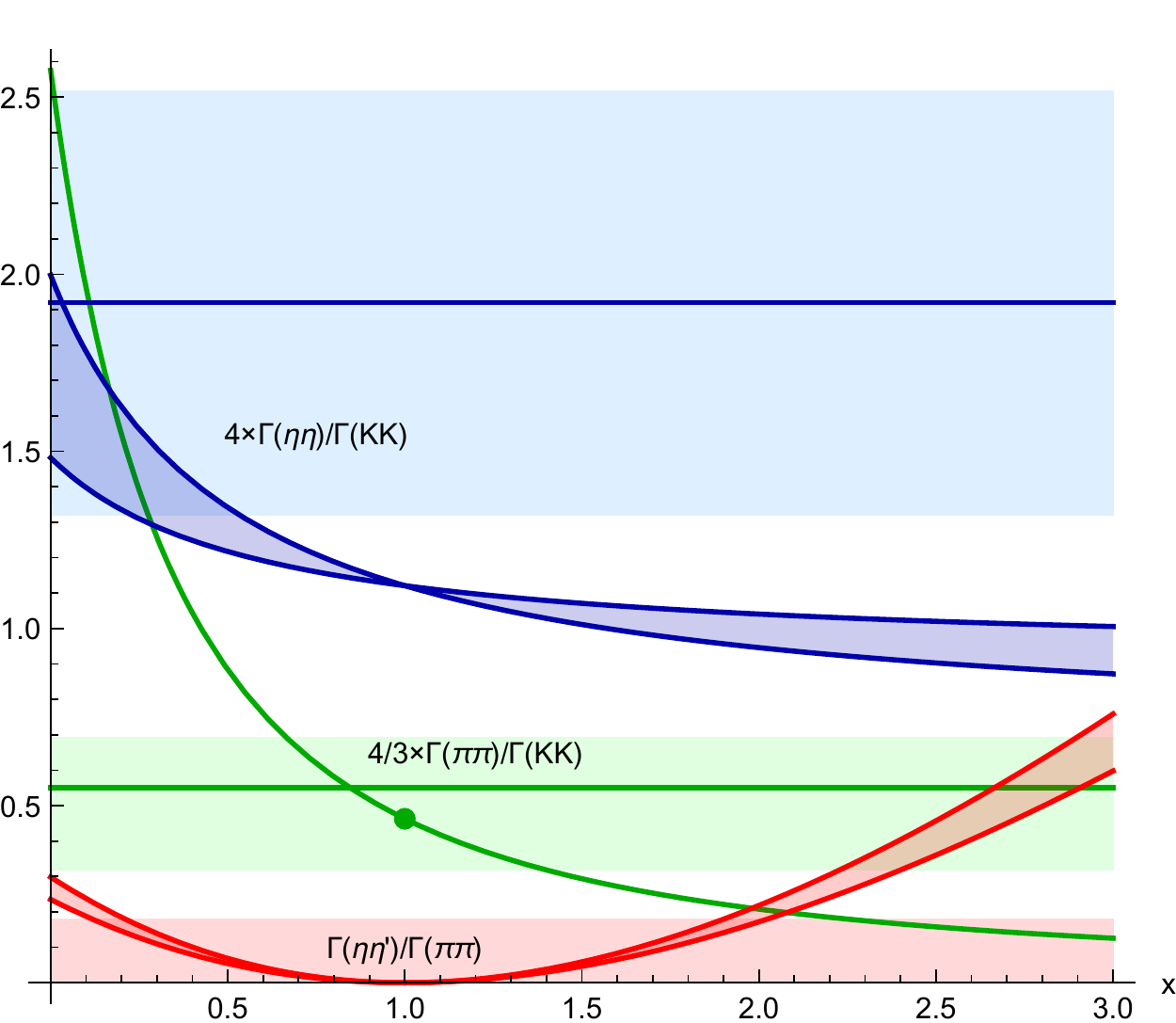}
\caption{Flavor asymmetries in the decay rates 
of the glueball $G_D$ of the Witten-Sakai-Sugimoto model with
mass set to that of $f_0(1710)$ as a function of $x=d_m/d_0$ and with $m_0$ varied over
the range implied by $\MKK=949$ MeV and $\lambda=12.55\ldots16.63$. 
The case $x=1$ studied in \BR\ is marked by a dot. 
The light-green and light-blue bands 
around the straight horizontal lines
give the current experimental results for
$f_0(1710)$ reported in \PDG, the light-red band corresponds to the upper limit on the $\eta\eta'$ decay rate
of WA102 \protect\cite{Barberis:2000cd}.
}
\label{fig1710}
\end{figure}

Fig.~\ref{fig1710} displays the holographic results for the ratios $(4/3)\times\Gamma(G_D\to\pi\pi)/\Gamma(G_D\to KK)$
(green curve) and $4\times\Gamma(G_D\to\eta\eta)/\Gamma(G_D\to KK)$ (blue curve), which describe the deviation from flavor symmetry in the decay
into two pseudoscalars, and the ratio $\Gamma(G_D\to\eta\eta')/\Gamma(G_D\to\pi\pi)$
when the mass of the glueball is set to that of the $f_0(1710)$ meson.
The rates involving the $\eta$ and $\eta'$ mesons depend on $m_0$, which is varied over the range 730 - 967 MeV
corresponding to the range of the 't Hooft coupling considered by us. These results are compared with the corresponding
experimental data and error bars reported by the Particle Data Group \PDG\ for $f_0(1710)$ (blue and green lines within light-blue and light-green bands) and the upper limit on the $\eta\eta'$ decay rate
reported by the WA102 experiment \cite{Barberis:2000cd}.

The value $x\approx0.32$ corresponds to the situation where kaon and pion decay rates have no mass dependence
other than through phase space factors; $x>0.32$ means nonchiral enhancement, $x<0.32$ the opposite.
The most symmetric choice $x=1$ considered by us in \BR\ is highlighted by the green dot in Fig.~\ref{fig1710}.
It is found to be within experimental error of the current experimental result \PDG. 
If one varies $x$ such the
result for $\Gamma(G_D\to\pi\pi)/\Gamma(G_D\to KK)$ does not leave this error band, one finds that
the Witten-Sakai-Sugimoto model with quark masses predicts
$\Gamma(G_D\to\eta\eta')/\Gamma(G_D\to\pi\pi)\lesssim 0.04$, which is well below the upper limit of 0.18 from WA102.
(This can be contrasted by the recent phenomenological study in \cite{Frere:2015xxa} which
predicts $\eta\eta'$ decay rates for $f_0(1710)$ that are several times higher than the upper limit reported by WA102.)

\begin{figure}
\includegraphics[width=0.45\textwidth]{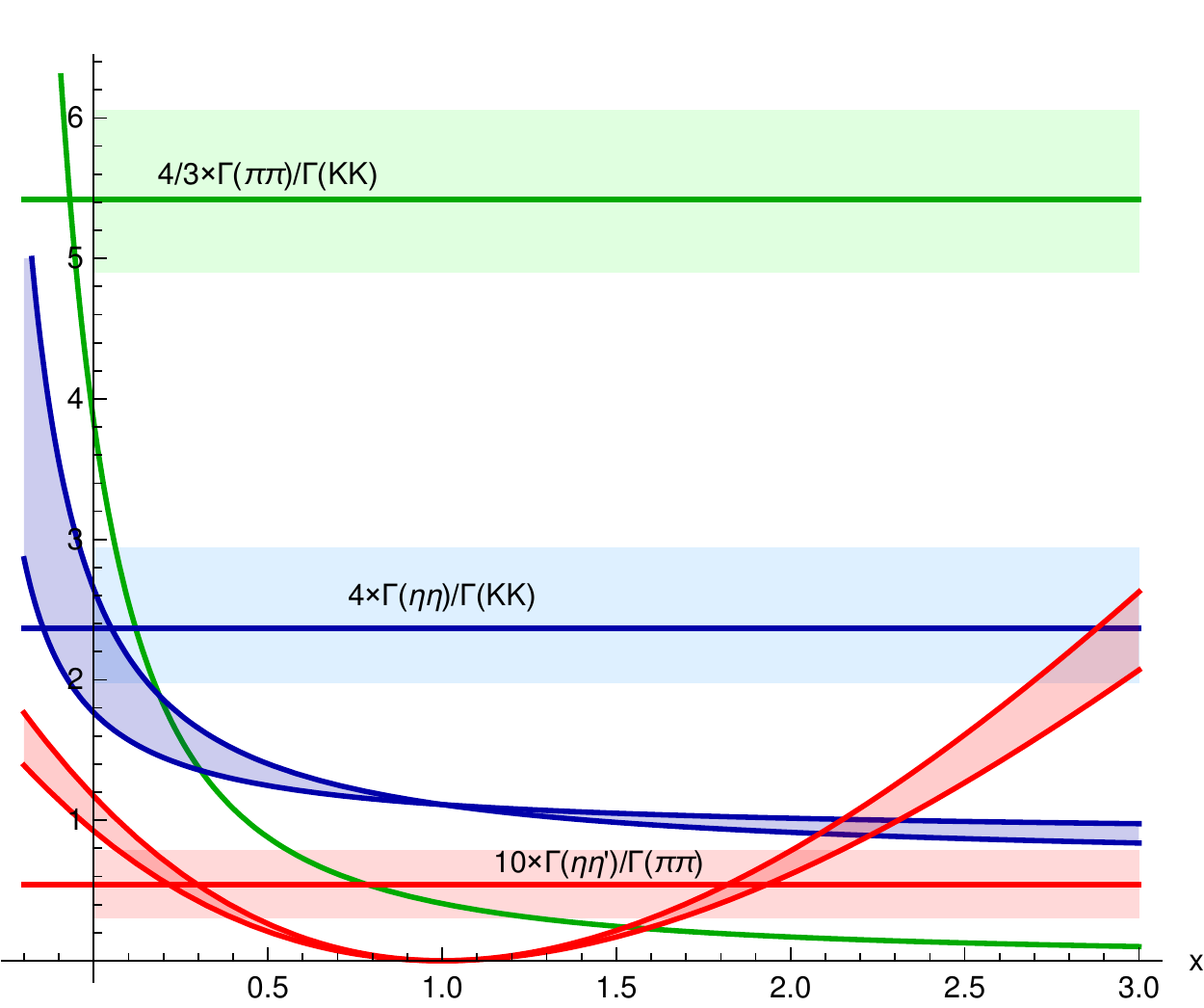}
\caption{Same as Fig.~\ref{fig1710}, but with glueball mass set to the mass of the $f_0(1500)$ meson.
For the latter, a nonzero decay rate into $\eta\eta'$ is reported by the Particle Data Group (red horizontal line and light-red band)
\PDG, here multiplied by a factor 10 for better visibility.}\label{fig1500}
\end{figure}

In Fig.~\ref{fig1500} the analogous comparison is made for a glueball with mass set to that of the $f_0(1500)$ meson.
Curiously enough, for $x\approx 0$, which corresponds in fact to a significant nonchiral suppression instead of enhancement,
the experimental results for the various ratios of decay into pairs of pseudoscalars are approximately reproduced.\footnote{In the case of $\Gamma(G_D\to\eta\eta')/\Gamma(G_D\to\pi\pi)$ a Breit-Wigner distribution for the glueball
mass was used in order to produce a nonzero result as the nominal mass of $f_0(1500)$ is below the $\eta\eta'$
threshold. (In the case of the $f_0(1710)$ this does not make much difference.)}
However, as mentioned in the Introduction, we have found this glueball candidate disfavored by the absolute value
of $\Gamma(G_D\to\pi\pi)$ (which depends only weakly on $x$) and even more so by $\Gamma(G_D\to4\pi)$, both
underestimating the experimental data significantly.

\section{Conclusion}

Based on the available experimental data on the decay rates for the glueball candidate $f_0(1710)$,
we conclude that the results of the Witten-Sakai-Sugimoto with finite quark masses %(where $\Gamma(G_D\to4\pi)$ remains
%approximately unchanged) 
are compatible with a nearly pure glueball interpretation of $f_0(1710)$, while disfavoring $f_0(1500)$.\footnote{%
In fact, whenever the $f_0(1500)$
is considered as the preferred glueball candidate, it is usually with significant mixing with $q\bar q$ states.}
The observed flavor asymmetries in the decay rates to pairs of pseudoscalars are remarkably well reproduced
for a range of the undetermined parameter $x=d_m/d_0$ over which the decay rate into $\eta\eta'$ remains well below
the current upper limit reported by WA102 \cite{Barberis:2000cd}. If the current experimental error \PDG\
on $\Gamma(G_D\to\pi\pi)/\Gamma(G_D\to KK)$ holds up, our holographic model predicts
$\Gamma(G_D\to\eta\eta')/\Gamma(G_D\to\pi\pi)\lesssim 0.04$.
Moreover, a ratio $\Gamma(G_D\to\eta\eta)/\Gamma(G_D\to KK)$ somewhat below the current experimental value
would be favored.

Additional predictions of the Witten-Sakai-Sugimoto model %where experimental data are still lacking
are that a pure glueball with a mass around 1.7 GeV
and thus above the $2\rho$ threshold should have
a substantial branching ratio into four pions \BPR: $\Gamma(G\to4\pi)/\Gamma(G\to2\pi)\approx 2.5$
(previous studies have usually assumed this to be negligibly small \cite{Close:2015rza})
and also into two $\omega$ mesons: $\Gamma(G\to2\omega)/\Gamma(G\to2\pi)\approx 1.1$. (Presently the $2\omega$
decay mode has the status of ``seen'' only, but
can be found to appear in radiative $J/\psi$ decays \PDG at the ratio of roughly $0.8(3)$ compared to the two pion rate.)

It would clearly be very interesting to see whether these predictions of the Witten-Sakai-Sugimoto model
for the decay pattern of $f_0(1710)$ (under the assumption of its nearly pure glueball nature\footnote{Even if mixing with $q\bar q$ states is indeed small, it may have nonnegligible effects on the final decay pattern. Absent a correspondingly improved holographic QCD model, one could consider a more phenomenological approach such as extended
linear sigma models \cite{Janowski:2014ppa} and use the holographic results
for the glueball-meson interactions as input instead of fixing those through
fits to experimental data.})
hold up against future experimental evidence.

\begin{acknowledgments}
We thank Denis Parganlija for collaboration on glueball decay in the massless Witten-Sakai-Sugimoto model,
and Jean-Marc G\'erard and Jean-Marie Fr\`ere for correspondence on $\eta$-$\eta'$ mixing.
This work was supported by the Austrian Science
Fund FWF, project no. P26366, and the FWF doctoral program
Particles \& Interactions, project no. W1252.
\end{acknowledgments}

\raggedright
%\bibliographystyle{kp}
%\bibliographystyle{h-physrev}
%\bibliographystyle{prsty}
%\bibliographystyle{apsrev4-1}
%\bibliographystyle{elsarticle-num}
%\bibliography{ar,tft,qft,books}
\bibliographystyle{JHEP}
\bibliography{glueballdecay}

\end{document}